%% file: valuation.tex
\documentclass[12pt,a4paper]{article}
\usepackage{epsf,array,delarray,amsmath,stmaryrd,amssymb}
\usepackage{graphicx}

\newtheorem{theorem}{Theorem}

\newcommand{\sA}{{\cal A}}

\newcommand{\sQ}{{\cal Q}}

\newcommand{\F}{{\cal F}}
\newcommand{\sT}{{\cal T}}
\newcommand{\sF}{{\cal F}}
\newcommand{\sO}{{\cal O}}

\newcommand{\R}{\mathbb R}
\newcommand{\nl}{\vskip 0.05 in \noindent}
\newcommand{\lb}{\llbracket}
\newcommand{\rb}{\rrbracket}
\newcommand{\bea}{\begin{eqnarray}}
\newcommand{\eea}{\end{eqnarray}}
\newcommand{\beast}{ \begin{eqnarray*} }
\newcommand{\eeast}{ \end{eqnarray*} }
\newcommand{\eoproof}{\hfill \makebox{\rule{0.8ex}{1.7ex}}\medskip}

\newcommand{\co}{ \hbox{\rm co}}
\newcommand{\tpi}{\tilde\pi}
\newcommand{\opi}{\,  ^0\Pi }

\newcommand{\be}{\begin{equation}}
\newcommand{\ee}{\end{equation}}

\begin{document}

\begin{center}
{\Large \bf Valuations and  dynamic convex risk measures}
\vskip 0.1 in
{\sc A. Jobert\footnote{
Statistical Laboratory, University of Cambridge, 
Wilberforce Road, Cambridge CB30WB, UK; A.Jobert@statslab.cam.ac.uk}
  \\and \\L.C.G. Rogers\footnote{
Statistical Laboratory, University of Cambridge, 
Wilberforce Road, Cambridge CB30WB, UK; L.C.G.Rogers@statslab.cam.ac.uk
The authors thank participants at the Isaac Newton Institute programme
Developments in Quantitative Finance 2005 for helpful discussions and
comments. We thank particularly Philippe Artzner, Alexander Cherny,
Phil Dybvig, Marco Frittelli, Lars Hansen, and Jose Scheinkman;
and two anonymous referees whose comments have resulted in 
numerous improvements.
}  
\\{\sl University of Cambridge}  }
%\begin{document}
%\maketitle
\vskip 0.2 in
First draft: March 2005; this version: August 2006.
\end{center}
\bibliographystyle{plain}

\begin{center}
%{\bf Preliminary and incomplete: comments welcome.}
\end{center}

\abstract{
This paper approaches the definition and properties of
dynamic convex risk measures through the notion of a family of concave
valuation operators satisfying certain simple and credible
axioms. Exploring these in the simplest context of a finite
time set and finite sample space, we find natural risk-transfer
and time-consistency properties for a firm seeking to spread
its risk across a group of subsidiaries.}

\input{S1.tex}

\input{S2.tex}

\input{S3.tex}

\input{S4.tex}

\input{S5.tex}

\input{conclusions.tex}

\newpage
\appendix
\input{appendix.tex}

\newpage
\bibliography{valuation}
%\bibliography{biblioriskmeasure}
%\bibliography{newbibrisk}

\end{document}

%% file: S1.tex
\section{Introduction.}

\bigskip
The growing literature of risk measurement considers mainly\footnote{
See \cite{ADEH,FS,delbaen,artzner,FS2,delbaen2,delbaen3,jaschke,
DepGer,Fri1,Fri3,cherny} for 
one-period risk-measurement, \cite{ADEHK} for the multiperiod
extension of \cite{ADEH}, \cite{riedel,CDK1,cvitanic,nakano} for a 
particular class of dynamic risk measures (the multiperiod behaviour 
addressed by the latter papers is somehow less general than the one considered
by \cite{ADEHK} and will be commented later),
\cite{Fri2,Fri4, Peng1, Peng2} for further dynamic risk measures,
and \cite{GS,ES} for 
related literature devoted to preference relations and 
Bayesian decision-making.  
} single-period risk measurement, where one attempts to
`measure' at time zero the risk involved in undertaking to receive
some contingent claim $X$ at time 1.
In this literature, 
a set $\sA$  of {\em acceptable} contingent claims is frequently
taken to be the primitive object (as in \cite{ADEH}, for example).
Such a set gives rise naturally to a {\em risk measure} $\rho_{\sA}$
via the definition
\[
\rho_{\sA}(X) = \inf \{ \, m\, |\,  X + m \in \sA\, \},
\]
which is simply the least amount of cash that would have to be added
to the contingent claim $X$ to make it acceptable.  A risk measure is naturally
{\em decreasing} in its argument; if added cash $b$ makes contingent claim $X$
acceptable, and if $X' \leq X$, then certainly $b+X'$ should be acceptable. 
This property makes the statement of various results rather clumsy and non-intuitive;
in common with others (for example, \cite{ADEHK}, \cite{CDK1},
\cite{CDK2}, \cite{Kloeppel_Schweizer} ), we shall instead speak of a
{\em valuation\footnote{This terminology is not standard, but is compact and
clear. A commonly-used term is `monetary utility function', which is 
descriptive if a little long-winded.}}, 
which is simply the {\em negative of a risk measure}. Thus in terms of 
the acceptance set $\sA$, we define the valuation $\pi_{\sA}$ by
\[
  \pi_{\sA}(X) \equiv -\rho_{\sA}(X) = \sup  \{ \, m\, |\,  X - m \in \sA\, \}.
\]

Expressed in this language,
Artzner et al. \cite{ADEH} define a {\em coherent valuation}
to be one which satisfies four  axioms equivalent to
\nl \nl \quad
(CV1)  \quad \textit{concavity}: $ \pi(\lambda X + (1-\lambda)Y)
 \geq \lambda \pi(X) + (1-\lambda) \pi(Y)  \quad (0\leq\lambda\leq 1)$;
\nl  \quad
(CV2)  \quad  \textit{positive homogeneity}: if \( \lambda \geq 0 \), then \(
  \pi(\lambda X) = \lambda \pi(X) \);
\nl  \quad
(CV3) \quad \textit{monotonicity}: if \( X \leq Y \), then \( \pi(X) \leq
  \pi(Y) \);
\nl  \quad
(CV4) \quad  \textit{translation invariance}: if \( m \in \mathbb{R} \), then \(
  \pi(Y+m) = \pi(Y) + m \).
\nl\nl
They go on to show that (under simplifying assumptions)
 any such valuation is representable as\footnote{
The properties (CV) appeared in an earlier paper of 
Gilboa \& Schmeidler \cite{GS}, in the context of Bayesian
decision theory. This study was not concerned with risk measurement.
}
\be
\pi(X) = \inf_{Q \in \sQ} E_Q [ X ],
\label{CohRM}
\ee
where $\sQ$ is some collection of probability measures\footnote{
Evidently, if $\pi$ has the form (\ref{CohRM}) then it satisfies the 
properties (CV1-4).
}. 

The positive-homogeneity  condition 
(CV2) is arguably unnatural, and was removed by F{\"o}llmer \&
Schied \cite{FS} and by Frittelli \& Gianin \cite{Fri1}
 who thereby introduced the notion of a {\em concave valuation}.
They show that a concave valuation admits a representation as
\be
\pi(X) = \inf_{Q \in \sQ} \bigl\lbrace \, E_Q[X] - \alpha(Q) \, \bigr\rbrace
\label{ConRM_rep}
\ee
where $\alpha$ is a concave `penalty' function on $\sQ$. Clearly if
$\alpha \equiv 0$, then we recover the representation of a coherent
valuation, but the notion of a concave valuation is more general.
\newline
\newline
Of course, the usefulness of a single-period study should be judged
by the extent to which it helps us to understand risk measurement in 
a multi-period setting; this has been well recognised for some time, and
recently attempts have been made to achieve that extension.  The
keywords `dynamic' and  `multi-period' occur frequently, but describe
very different notions.  One of these is where the goal is
to value at intermediate times some contingent claim to be received
at the terminal time $T$; this is in some sense an interpolation of 
valuations, which nevertheless must be done in a naturally consistent
way.  Examples of this kind of study include Peng \cite{Peng2}, 
Detlefsen \& Scandolo \cite{DetScan},
Kl\"oppel \& Schweizer \cite{Kloeppel_Schweizer},
Cheridito \& Kupper  \cite{Cheridito_Kupper}, F\"ollmer 
\& Penner \cite{FP}. Another notion
of a dynamic measurement of risk is to take some random cashflow,
and ascribe some value to it at time 0:  the contributions of 
Artzner, Delbaen, Eber, Heath and Ku \cite{ADEHK},
F{\"o}llmer and Schied \cite{FS,FS2},
 Cvitani{\`c} and Karatzas
\cite{cvitanic}, Nakano \cite{nakano}, Cheridito, Delbaen
and Kupper \cite{CDK1}, and Riedel \cite{riedel}
are of this type.
The notion of `dynamic' risk measurement which we plan to study
in this paper takes {\em random cash balance} processes as the 
inputs, and returns random processes, the {\em valuations as functions
of time}, as the output. This seems to us to be the setting in which
one would want to apply ideas of risk measurement. Moreover,
the dynamic fluctuation of the cash balance is clearly the essence 
of cashflow problems, and therefore
 of risk measurement; it is {\em not} sufficient to consider only the
total amount of cash accumulated by some arbitrary time in the future,
as the experience of Long Term Capital Management demonstrates.
This (fullest) notion of dynamic risk measurement is as yet little
studied: Scandolo \cite{Scandolo_2003},\cite{Scandolo_2004}, Frittelli \& 
Scandolo \cite{Fri4},  and Cheridito, Delbaen
\& Kupper \cite{CDK2} are contributions of this type.

%For example,
 %Artzner, Delbaen, Eber, Heath and Ku \cite{ADEHK} adapt the static
%coherent valuation axioms to the product sample space $\{0,1,\ldots,T \}
%\times \Omega$
%and  obtain a representation for a coherent valuation of the form:
%\begin{displaymath}
%\pi(X)
%= \inf_{V \in \sV } E_P \left[
  %\sum_{t=0}^T  X_t (V_t - V_{t-1}) \right]
%\end{displaymath}
%for one fixed probability measure $P$  and a set $\sV$
%of positive increasing adapted processes. The argument of $\pi$ is of course
%a {\em cash-flow} process.
%\newline
%\newline
%F{\"o}llmer and Schied \cite{FS,FS2}, Cvitani{\`c} and Karatzas 
%\cite{cvitanic} and Nakano \cite{nakano} also have several periods for 
%portfolio construction and measure the risk of final wealths, which may 
%depend on trajectories. However, they do not incorporate in their analysis 
%what happens at intermediate dates. The trajectories have only an impact 
%on the final values.
 %Similarly, in Riedel \cite{riedel}, the risk-adjusted measurement of 
%cumulative cash-flows involves only final values and not the whole trajectories. 
%By contrast, as in \cite{ADEHK}, we  not only look at
%final values but also at intermediate time points. Although Cheridito, Delbaen
%and Kupper \cite{CDK1} consider continuous-time discounted value
%processes, the convex risk measures treated in their paper are static as they
%only measure the risk of a discounted value process at the beginning of a
%given time period.

The major difference between the static and  multi-period frameworks
 is the issue of
{\em dynamic consistency}. Although every set of probability measures generates a
 coherent valuation in the static framework, only
sets of probability measures consistent in an appropriate sense
 yield dynamic coherent valuations.
 This consistency property of
probability measures (or stability by ``pasting'') has been analysed by 
Epstein and Schneider \cite{ES} (building upon the atemporal multiple-priors model 
of Gilboa and Schmeidler \cite{GS} and using prior-by-prior Bayesian updating
 for ``rectangular'' sets of priors), Artzner et al. \cite{ADEHK} 
(using change-of-measure martingales) and Riedel \cite{riedel} (via Bayesian 
updating and a different kind of translation invariance property). It is 
often referred to as \textit{multiplicative stability} \cite{delbaen}.
The axiomatic approach of this paper has also been independently proposed
by Cheridito, Delbaen \& Kupper \cite{CDK2}. Their study thoroughly
explores the implications of the given setup for acceptance sets and
coherent risk measures, relating to earlier work of the authors, 
providing (Theorem 4.6) a nice characterisation of how the acceptance
sets combine intertemporally. 
We have nothing to add to the understanding of the 
acceptance sets, because the emphasis here is quite different;
we take the valuations themselves as fundamental (rather than the acceptance
sets), and we aim to discover what  consequences of the
axiomatic setup can be developed. 
\newline
\newline
In this  paper, we present and analyse\footnote{
We work mainly in the technically simple setting of a finite time set, 
and a finite probability space $\Omega$; this allows us to obtain
the main ideas without being held up by technical issues.
} the notion of a {\em dynamic family of 
 concave valuations},  extending the dynamic coherent
valuation of \cite{ADEHK}, rather as 
Frittelli \& Gianin \cite{Fri1} and  F{\"o}llmer \& Schied \cite{FS}
extend \cite{ADEH} in the single-period context.

The basic object of study here is a family of {\em valuations}.
To see why such a starting point may be useful, we quote a
simple result which is presumably well known (it certainly appears
in Rogers \cite{R1}, for example.) The idea is to write down certain
natural axioms that {\em market valuation operators} should have, and to 
derive implications\footnote{Peng \cite{Peng2} develops a set of axioms
for non-linear valuations which are similar in some respects. For example,
if we take his axioms (A1)--(A4) and assume positivity and linearity as
well, then we obtain the axiom (A3) of Theorem \ref{FTAP}.
}.

%%%%%%%%%%%%%%%%%%%%%%%%%%%%%%%%%%%%%%%%%%%%%%%%%%%%%%%%%%%%%%%%%%%%%%
\begin{theorem}\label{FTAP}
In a filtered probability
space $(\Omega, ( {\cal F}_t )_{t \geq 0}, P)$, suppose that
valuation operators $(\pi_{tT})_{0\leq t \leq T}$
\[
\pi_{st} : L^\infty(\F_t) \rightarrow L^\infty(\F_s)
        \quad\quad (0 \leq s \leq t).
\]
satisfy the following four axioms:\vskip 0.05in \noindent
(A1)\quad
 Each $\pi_{st}$ is a bounded positive linear operator from $L^{\infty}(\F_t)$
to $L^{\infty}(\F_s)$;
\vskip 0.05in \noindent
(A2) \quad
 If $Y \in L^\infty(\F_t)$, $Y \geq  0$, then
\[
         \pi_{0t}(Y) = 0 \iff P(Y>0)= 0.
\]
({\em no arbitrage})
\vskip 0.05in \noindent
(A3)\quad For $ 0 \leq s \leq t \leq u$, $Y \in L^\infty(\F_u)$,
$X \in L^\infty(\F_t)$,
\[
\pi_{su}(XY) = \pi_{st}( X \pi_{tu}(Y) )
\]
({\em dynamic consistency})
\vskip 0.05in \noindent
(A4) \quad
If $(Y_n) \in L^\infty(\F_t)$, $|Y_n| \leq 1$,
$Y_n \uparrow Y$ then $\pi_{st}(Y_n) \uparrow \pi_{st}(Y)$
({\em continuity})
\newline

For simplicity, suppose also that ${\cal F}_0$ is trivial.
Then there exists a strictly positive process
$(\zeta_t)_{t \geq 0}$ such that the valuation operators $\pi_{st}$ can be
expressed as
\be        \pi_{st}(Y) = \frac{ E \bigl[ \, \zeta_t Y \,\bigl\vert \, {\cal F}_s \,
						 \bigr]}
                {\zeta_s} \quad  (0 \leq s \leq t).
\label{price}
\ee

\end{theorem}
%%%%%%%%%%%%%%%%%%%%%%%%%%%%%%%%%%%%%%%%%%%%%%%%%%%%%%%%%%%%%%%%%%%

The proof of this result takes about a page, and is included in the appendix;
nothing more sophisticated than standard facts about measure theory is
required\footnote{
Note however that the assumption that the valuation operators are
defined on the {\em whole} of $L^\infty(\sF_t)$ greatly simplifies the argument.
}.  However, its importance is not to be underestimated; it is
in some sense a substitute for the
{\em Fundamental Theorem of Asset Pricing} (FTAP). 
Indeed, the  FTAP implies a risk-neutral valuation
principle (\ref{price}),
 but its axiomatic starting point is different; in the FTAP we start
from some suitably-formulated
 axiom of absence of arbitrage, and here we start from the axioms
(A1)--(A4). Which of these two axiomatic starting points one should wish
to assume is of course a matter of taste; in defence of the unconventional
approach taken here, it is worth pointing out\footnote{
It is also worth pointing out that it took years to find the correct
formulation for the notion of absence of arbitrage!
} that if we want to have the valuation principle
 (\ref{price}) for all $Y \in L^\infty(\sF_t)$, 
then (A1)--(A4) must hold anyway!

Of the four axioms assumed in Theorem \ref{FTAP}, the key one is the
dynamic consistency axiom, (A3), as you will see from the proof;
without this, we are able to prove that (\ref{price}) holds if $s=0$, but
this is of course far too limited to be useful. Notice the interpretation
of (A3); we can obtain $X$ units of $Y$ at time $u$ in two ways, either
by buying at time $s$ the contingent claim $XY$, or by buying at time $s$
the contingent claim which at time $t$ will deliver $X$ units of the
time-$t$ price $\pi_{tu}(Y)$ of $Y$, and (A3) says that these two 
should be valued the same at time $s$.

Now Theorem \ref{FTAP} relates to {\em market} valuations, where
linearity in the contingent claim being priced is a reasonable assumption;
if we want to buy $X$ and $Y$, the price will be the price of $X$ plus the 
price of $Y$. However, when it comes to risk measurement, what the valuation
is doing is to tell us how much capital a given firm should set aside
to allow it to accept a named cash balance. Linearity now would {\em not}
be a property that we want (we might require a positive premium both to 
cover a cash balance $C$ and to cover $-C$, but we would not require a 
positive premium to cover the sum of these). Moreover, the valuations
 will depend on the particular firm; different firms will
have different valuations, and an interesting question is how 
these combine.

In the next Section, we shall formulate the analogues of the axioms of
Theorem \ref{FTAP} for {\em concave valuations}, and deduce some
of their consequences. There are substantial differences; concave valuations
 have to be defined over cash balances, because without linearity
we cannot build the price of a cash balance from the prices of its component
parts. Nevertheless, the dynamic consistency axiom turns out to
be the heart of the matter. We shall characterise families of valuations
which satisfy the given axioms; it turns out that
such families (and their duals) possess simple and appealing 
recursive structure.

We shall also study the question of how a firm may decide to divide
up a risky cash balance process between its subsidiaries, each of which
is subject to the regulatory constraints implicit in their individual
valuations. We find that there is an optimal way to do this
risk transfer, in terms of a sup-convolution (as in, for example,
 the study of Barrieu
and El Karoui \cite{BEK}, Kl\"oppel \& Schweizer
\cite{Kloeppel_Schweizer}.) Moreover, the optimal risk transfer generates
a family of valuations for the firm as a whole, and this
family of valuations satisfies {\em the same axioms as the 
individual components}.  We shall also see that if the firm decides
at time 0 how it is going to divide up the cash balance between
its subsidiaries, then at any later time, whatever has happened in 
the meantime, the original risk transfer chosen is still optimal. There
is therefore a time-consistency in how the firm should transfer
risk among its subsidiaries.

Another question we answer concerns what happens if a firm facing 
a risky cash balance process is allowed to take offsetting positions
in a financial market. We find that there is an optimal offsetting position
to be taken, which is time consistent, and the induced valuations
for the firm once again satisfy the axioms.

%% file: S2.tex
%%%%%%%%%%%%%%%%%%%%%%%%%%%%%%%%%%%%%%%%%%%%%%%%%%%%%%%%%%%%%%%%%%%%%%%%%%
\section{Dynamic concave valuations.}\label{S2}
Working in a filtered probability space $(\Omega, \sF, (\sF_t)_{0
\leq t \leq T}, P)$, we let $BV$ denote the space of adapted processes of 
bounded variation with $R$-paths\footnote{
This is the terminology of Rogers \& Williams
\cite{RW} for paths that are right continuous
with left limits everywhere.
}.  We think of $K \in BV$ as a {\em cash balance process}, with $K_t$
being interpreted as the total amount of cash accumulated by time $t$.
The process $K$ need not of course be increasing. The upper end $T$ of
the time interval considered is a finite constant; there is no real
difficulty in letting the time set be $[0,\infty)$, but we choose not
to do this here in view of our concentration later on examples where
$\Omega$ is finite.  

We propose to introduce some natural axioms to be satisfied by
a family \footnote{
As usual, $\sO$ denotes the optional $\sigma$-field on $[0,T]
\times \Omega$,  and by extension the statement
$\tau \in \sO$ for a random time $\tau$ means that $I_{ [\tau,T]}$
is an optional process, equivalently, that $\tau$ is a stopping time
- see, for example, \cite{RW}  for more background on the general theory of
processes.
}
\[
\{ \pi_\tau: BV \rightarrow L^\infty(\sF_\tau) \;
\bigl\vert\; \tau \in \sO
\}
\]
of   valuations.
%----------------------------------------------------------------
We interpret 
$- \pi_\tau(K)$ as the amount of capital required by law at time
$\tau$ to allow a firm to accept the cash balance process $K$.
The requirement could be different for firms in different
countries, or for an investment bank and a hedge fund, for 
example.
%Not surprisingly, we shall suppose that
%\be
	%\pi_\tau(0) = 0 \quad \forall \tau.
%\label{zero_level}
%\ee

\noindent
The axioms we
require of the family of valuations are the following.
\begin{center}%%%%%%%%%%%%%%%%%%%%%%%%%%%%%%%%%%%%%%%%%%%%%
\setlength{\fboxrule}{1pt}
\setlength{\fboxsep}{5mm}
\fbox{
\parbox[c]{0.9\linewidth}{
\nl
(C) $\pi_\tau$ is concave for all $\tau$;
\nl
(L)  $\pi_\tau(I_AI_{[\tau,T]}K)=I_A\pi_\tau(K)$ for all $\tau, K$, for all
$A \in \sF_\tau$;
\nl
(CL) %if $\tau$, $\tau'$ are two stopping times, and $A \in \sF_\tau \cap
%\sF_{\tau'}$, with $A \subseteq \{ \tau = \tau'\}$, then 
%for every $K$
%\[
%\pi_\tau(K) = \pi_{\tau'}(K) \quad \hbox{\rm on $A$}
%\]
if $\tau$, $\tau'$ are two stopping times, then
\[
      \pi_\tau(K) = \pi_{\tau'}(K) \quad \hbox{\rm on $\{\tau = \tau'\}$ };
\]
\nl
(M) if $K_t \geq K'_t$ for all $t$, then $\pi_\tau(K) \geq\pi_\tau(K')$
       for all $\tau$;
\nl
(DC) for stopping times $\tau \leq \sigma$, 
\[
    \pi_\tau(K) = \pi_\tau(KI_{[\tau,\sigma)} + \pi_\sigma(K)
             I_{[\sigma, T]});
\]
\nl
(TI)  if for some $a\in L^\infty(\sF_\tau)$
 we have  $K_t = K'_t+a$ for all $t \geq \tau$, 
then $\pi_\tau(K)=a+\pi_\tau(K')$;
\nl
(Z) $\pi_\tau(0) = 0 \quad \forall \tau \in \sO$.
}
}
\end{center}%%%%%%%%%%%%%%%%%%%%%%%%%%%%%%%%%%%%%%%%%%%%%%%%

\nl
{\sc Remarks.} Axiom (C) is a natural property for capital
adequacy requirements for risky cash balances; see \cite{ADEH},
for example.
%Indeed, if $\preceq$ denotes a 
%convex monotone preference relation\footnote{
%We require (i) $K_t \leq K_t' \Rightarrow K\preceq K'$ ;
%(ii) $\{ K: K \succeq K'\}$ is convex for each $K'$;
%(iii) $\preceq$ is transitive.
%We write $K \sim K'$ if $K \preceq K'$ and $K' \preceq K$.
%} on cash balances, then the reservation buy price for $K$ is that
%scalar $\pi(K)$ such that $K-\pi(K) \sim 0$. Concavity of $\pi$ is
%an immediate consequence.

Axiom (L) (for {\em local})
says two things. Firstly, if you have reached time $\tau$, then 
all that matters for valuation is how much cash has {\em currently}
been accumulated, and what is to come;  the exact timing of the earlier
payments does not influence the valuation\footnote{
Note that Axiom (L) does {\em not} say that you value the
increments of the 
cash balance $K$  after $\tau$ the same as
the whole of the original cash balance $K$! The
axiom says that you value $K$ the same as the
cash balance $I_{[\tau,T]}K$, which
 pays nothing up til time $\tau$, then a lump
sum of $K_\tau$.
}.  Secondly, Axiom (L) expresses the following natural fact:
 at time $\tau$,
if event $A$ has not happened then the cash balance $I_AI{[0,\tau)} K$ is 
clearly worthless,
and if the event has happened, then the cash balance $I_AI{[0,\tau)}K$ will 
be worth 
the same as $K$.  It is easily seen that (L) implies the following
useful consequences:
\beast
   \pi_\tau(K) &=& \pi_\tau( I_{ [\tau,T] } K ),
\\
   \pi_\tau(I_A K ) &=& I_A \pi_\tau(K) 
\\
             &=& \pi_\tau(I_AI_{ [\tau,T] } K )
\\	
	&=& I_A \pi_\tau( I_A K )
\eeast
for any $\tau \in \sO$, $A \in \sF_\tau$, and cash balance $K$.
Another useful consequence of (L) is the property

\nl
\begin{center}
(Z)\quad\quad  $\pi_\tau(0) = 0 \quad \forall \tau \in \sO$,
\end{center}
\nl 
which we see by taking $A = \emptyset$.

Axiom (CL) (for {\em consistent localisation})
says that the localisations of $\pi_\tau$
and $\pi_{\tau'}$ agree where $\tau = \tau'$, again a natural condition.

Axiom (M) (for {\em monotonicity}) says that a larger capital
reserve is required to short a larger cash balance,
 but it says more than just this. In particular, if $K_t \geq K'_t$
for all $0\leq t\leq T$, with $K_T = K'_T$, then the two cash balances $K$
and $K'$ both deliver exactly the same in total, but $K$ is considered less
risky than
$K'$ because it delivers the cash {\em sooner}.  An earlier version of this
work used an axiom which expressed indifference between cash balances that
delivered the same total amount of cash; though this axiom was entirely
workable, the effect of it was that the valuations were essentially
defined on cash balances which were all delivered at time $T$, and the 
valuations themselves served only to `interpolate' prices in some
sense.  The interpretation of (M) is not that earlier payments are
preferred to later payments because of the {\em interest} that will accrue;
indeed, we think of all payments as being discounted back to time-0
values (or equivalently that the interest rate is zero). Even under
these assumptions, according to (M)  earlier payments  are better than
later ones - as in reality they are!  This embodies the essence of cashflow
problems, where a firm may be in difficulties not because it does
not have sufficient money owed to it, but because that money has not
yet come in.

Axiom (DC) (for {\em dynamic consistency}) has a simple and natural 
interpretation. It says that we must set aside as much for  the cash balance
$K$, as for the cash balance which gives us $K$ up to time $\sigma$, and
at time $\sigma$ requires us to hand in the accumulated cash balance
$K_\sigma$ in return for the amount of cash that we would allow us to accept the
entire cash balance $K$.  This latter cash balance would clearly 
allow us to accept the original cash balance $K$. It is worth remarking
that in some other studies the expression of the notion of dynamic consistency
appears much simpler: see, for example, \cite{Peng2}, \cite{DetScan}, where it
is possible to express dynamic consistency as $\pi_\tau = \pi_\tau 
 \pi_\sigma$ for any stopping times $\tau \leq \sigma$. However,
do note that these studies are only concerned with valuing {\em terminal}
cash balances; if we restrict the condition (DC)
to cash balance processes which are non-zero only at time $T$, then
we get this same simple form.  When valuing only 
terminal cash balances, the intermediate valuations are simply numbers
 which do
not correspond to any cash value. By contrast, in the setting we
are using, the intermediate valuations {\em must} be denominated in 
cash (or some other asset), because we are going to have to consider
exchanging a future cash balance process for cash today. This is why
the axiom (DC) looks a little more involved\footnote{
This also motivates the common use of the clumsier term `monetary
utility function' instead of `valuation'.
}. \cite{CDK2} use exactly
the same criterion.

The next axiom (TI) (for {\em translation invariance}) is again
entirely natural if we think (as we do) that valuations should have
a monetary interpretation. We hope later to see what can be done if
(TI) is abandoned, as this leads us back closer to the idea of recursive
utility (\cite{EZ}, \cite{DE}, \cite{Sk});
 however, the first thing that will need to be done is to 
revise the notion of dynamic consistency.

Finally, axiom (Z) (for {\em zero level}) is again a natural 
consequence of the notion that $\pi_\tau(K)$ is the capital reserve
required to allow the firm to accept cash balance process $K$.

In the next Section, we shall
 explore the consequences of these axioms only in the simplest
possible setting, where $\Omega$ is {\em finite}. This means in particular
that we can take the time set to be finite, and the entire filtered
probability space to be represented by a tree.  This (restrictive)
assumption allows us to ignore all technicalities, and quickly
uncover  essential structure implied by the axioms. We
 remark only that in any real-world
application we would be forced to use a numerical approach, in which
case we would have to be working with a finite sample-space.

%% file: S3.tex
%%%%%%%%%%%%%%%%%%%%%%%%%%%%%%%%%%%%%%%%%%%%%%%%%%%%%%%%%%%%%%%%%%%%%%%%%%
\section{Valuations on finite trees.}\label{S3}
Henceforth, we work with a finite sample space $\Omega$, and a
finite time set $\{0,1,\ldots,T\}$. The  $\sigma$-field $\F$ on 
$\Omega$ is of course the $\sigma$-field of all subsets, and the
filtration is
represented by a  tree\footnote{
The tree does not of course have to be binomial, or regular.
} with vertex set $\sT$. The root of the tree
will be denoted by 0, and from any vertex $y \in \sT$ there is a
unique path to 0; we shall say that $y$ is a descendant of $x$ 
(written $x \preceq y$) if $x$ lies on the path from $y$ to 0.
If $x \in \sT$, we shall write $x-1$ for the immediate ancestor
of $x$, $x+1$ for the set of immediate descendants of $x$, and
$x+$ for the set of all descendants of $x$, including
$x$ itself. Note that $\Omega$ can be identified with the set of
endpoints of $\sT$. 
For any $x \in \sT$ we shall
denote by $t(x)$ the {\em time} of $x$, which is the depth of 
$x$ in the tree. Thus $t(0)=0$; $t(y) = t(x)+1$ for
any $y \in x+1$; and $t(x) = T$ for any 
terminal node $x$. Notice also that a stopping time
$\tau$ can be identified with a subset\footnote{
The {\em graph} of $\tau$ - see \cite{RW}.
} $\lb \tau \rb$ of $\sT$ with the property that for any terminal
node $\omega$ of the tree the unique path from $\omega$ to 0 intersects
$\lb \tau \rb$ in exactly one place.

In this setting, a {\em cash balance} is simply a map $K : \sT \rightarrow \R$.
We interpret $K_x$ as the cumulative amount of the cash balance at
vertex $x$ in the tree.
We shall  also suppose throughout that interest rates are zero,
or equivalently that cash balances have all been discounted 
back to time-0 values; this assumption is insubstantial, and 
leaves us clear to focus on what is important here.

In view of axiom (C), the valuations $\pi_\tau$ are
just concave functions defined on some finite-dimensional
Euclidean space, and so can be studied through their convex
dual functions
\[
{\tilde\pi}_\tau(\lambda) \equiv \sup_K \{ \pi_\tau(K)
		- \lambda \cdot K \}.
\]
For simplicity of exposition, we shall make the assumption
\nl\nl

%\begin{center}
\noindent
{\sc ASSUMPTION A:} {\sl For every $\tau$, the valuation $\pi_\tau$
is concave, strictly increasing,
and
$C^2$  in the relative interior of its domain of finiteness.}
%\end{center}
\nl\nl
\noindent
%From the final condition it can be easily shown
%that if $\pi_y(K) = -\infty$ and $x \preceq y$,
%then also $\pi_x(K) = -\infty$.

\nl\nl
By duality, the original functions $\pi_\tau$ can be expressed as
\be
	\pi_\tau(K) = \inf_{\lambda} \{ \lambda \cdot K +
		 {\tilde\pi}_\tau(\lambda) \};
\label{convex_dual}
\ee
compare with the equation (\ref{ConRM_rep}) above, as in  F\"ollmer \& Schied
\cite{FS}, Frittelli \& Gianin \cite{Fri1}.
That equation is at one level simply the general statement (\ref{convex_dual})
of duality, but with a bit more; in (\ref{ConRM_rep}) the infimum is
taken over a family of probability measures, and in (\ref{convex_dual})
the infimum is unrestricted. We shall later see that the axioms
used here do in fact imply that $\lambda$
 must be a probability on $x+$ for each $x \in \lb \tau \rb$.

\subsection{Decomposition.}%%%%%%%%%%%%%%%%%%%%%%%%%%%%%%%%
\noindent
The dynamic consistency axiom (DC) and localisation axioms (L), (CL)
 allow us to decompose the valuations in a simple way. To see
this, notice firstly that the family $(\pi_\tau)$ of valuations
is determined once the smaller family $\{ \pi_x: x \in \sT \}$ is
known, where for $x \in \sT$ the operator $\pi_x$ is defined to be
\be
	\pi_x  = \pi_{\tau_x},
\label{pi_x}
\ee
where $\tau_x$ is the stopping time
\bea
	\tau_x(\omega) &=& t(x) \quad\hbox{\rm if $x \prec \omega$;}
\label{tau_x_def}
\\
		&=& T \quad \hbox{\rm otherwise}.
\nonumber
\eea
Here, of course, we are identifying $\Omega$ with the set of terminal nodes of 
$\sT$.  Once we know the operators $\{ \pi_x: x \in \sT \}$, Axioms (L)
and (CL) allow us to put together any of the $\pi_\tau$.

However, the $\pi_x$ can themselves be assembled from the family 
$\{ \pi_{x,x+1}:  x \in \sT   \}$ of {\em one-step valuations}, 
defined in the following way. If $x$ is a terminal node, then the argument
of $\pi_{x,x+1}$ is  a cash balance $k$ defined at $x$, and $\pi_{x,x+1}(k)
=\pi_x(k)$. For all other $x$, given a cash balance $k$ defined on $x \cup x+1$, 
we extend this to a cash balance $\bar k$ defined on all of $\sT$ by
\beast
	\bar k_z &=& k_x \quad \hbox{\rm if $z=x$;}
\\
	         &=& k_y \quad \hbox{\rm if $y \preceq z$ for some $y \in x+1$;}
\\
		&=& 0 \quad \hbox{\rm otherwise}.
\eeast
We may then define 
\be
	\pi_{x,x+1} (k) = \pi_{\tau_x}(\bar k).
\label{pi_x_x+1}
\ee
Of course, the point of this decomposition is really the {\em converse}:
we wish to build the (complicated) family $(\pi_\tau)_{\tau \in \sO}$
from the simpler family $(\pi_{x,x+1})_{x \in \sT}$ of one-step
valuations. It is clear that if we derive $(\pi_{x,x+1})_{x \in \sT}$
from a family $(\pi_\tau)_{\tau \in \sO}$ satisfying the axioms
given in Section \ref{S2}, then the family of one-step  valuations
must have the following properties:
\begin{center}%%%%%%%%%%%%%%%%%%%%%%%%%%%%%%%%%%%%%%%%%%%%%
\setlength{\fboxrule}{1pt}
\setlength{\fboxsep}{5mm}
\fbox{
\parbox[c]{0.85\linewidth}{
\nl
{\em (c)} $\pi_{x,x+1}$ is concave;
\nl
{\em (m)} if $k_z \geq k'_z$ for all $z \in x \cup x+1$ then 
        $\pi_{x,x+1}(k) \geq \pi_{x,x+1}(k')$;
\nl
{\em (ti)}  if $k_z = k'_z+a$ for all $z \in x \cup x+1$, 
 then 
         $\pi_{x,x+1}(k) = \pi_{x,x+1}(k')+a$;
\nl
{\em (z)} $\pi_{x,x+1}(0) = 0$.
}
}
\end{center}%%%%%%%%%%%%%%%%%%%%%%%%%%%%%%%%%%%%%%%%%%%%%%%%
%as well as the property $\pi_{x,x+1}(k) = k_x$ if $x$ is a terminal node.
What we now argue is that given a family $(\pi_{x,x+1})_{x \in \sT}$ of one-step
valuations satisfying {\em (c), (m), (ti), (z)} we can build a family
$(\pi_\tau)_{\tau \in \sO}$ of valuations satisfying the axioms
of Section \ref{S2}.

The essence of the construction is to get the $(\pi_x)_{x \in \sT}$, for then if 
we have a stopping time $\tau$ we define
\[
	\pi_\tau(K) = \pi_z(K) \quad \hbox{ at $z \in \lb\tau\rb$}.
\]
To get the $(\pi_x)_{x \in \sT}$, we proceed by backward induction,
assuming that we have constructed $\pi_x$ for all $x$ such that
$t(x) \geq n$. The induction starts, because if $x$ is a terminal
node we have $\pi_x(k) = \pi_{x,x+1}(k)$, and if $t(x) = n-1$ we may define
\[
	\pi_x(K) = \pi_{x,x+1}(k),
\]
where $k$ is the cash balance defined by
\beast
	k_z &=& K_x \quad\quad\hbox{\rm if $z=x$;}
\\
	&=& \pi_z(K) \quad\hbox{\rm if $z \in x+1$.}
\eeast
There is no problem with this, as the definition of $\pi_x$ requires
only the one-step operator $\pi_{x,x+1}$ and the operators 
$(\pi_z)_{z \in x+1}$ which are already known (by the inductive hypothesis).

If we vary the notation for $\pi_{x,x+1}(k) \equiv \pi_{x,x+1}(k_x,
k_{x+1})$ so as to make the dependence on the cash balances at node $x$
and nodes $x+1$ explicit, then the recursive construction of the
$\pi_x$ takes the clean form
\be
	\pi_x(K) = \pi_{x,x+1}(K_x, \pi_{x+1}(K) ).
\label{primal_recursion}
\ee
Notice the formal similarity to the notion of {\em recursive 
utility} - see Epstein \& Zin \cite{EZ}, Duffie \& Epstein
\cite{DE}, Skiadas \cite{Sk}.  This similarity is 
only formal; the theory of recursive utility deals with
preferences over running consumption processes, 
a notion that does not feature in 
our present discussion. This is an interesting possible variant
of the axiomatic approach which we hope to return to
at a later date.

It remains to see that the operators $(\pi_\tau)_{\tau \in \sO}$
defined by (\ref{primal_recursion})
satisfy the axioms given in Section \ref{S2}.

Property (C) follows from the concavity property {\em (c)} by backward
induction.  Properties (L) and (CL) are immediate from the construction.
Property (M) follows from {\em (m)}, again by backward induction.
Property (DC) requires a little more thought (and use of the property
{\em (ti)}), but again follows from the construction. Finally,
property (TI) is a consequence of {\em (ti)}.

\subsection{Duality.}%%%%%%%%%%%%%%%%%%%%%%%%%%%%%%%%%%%%%%%%%%%%%%%%
We have just seen that the axioms permit us to decompose the valuations
 into simpler pieces, but what is the corresponding result for
the dual valuations $\tilde\pi_x$?  What are the characteristic
properties? 

To understand the structure of the dual,
 firstly note that the dual valuation
\be
{\tilde\pi}_x(\lambda) \equiv \sup_K \{ \pi_x(K)
                - \lambda \cdot K \}
\label{pi_tilde_def}
\ee
is not always going to be finite. Indeed, because of (L) and (CL),
${\tilde\pi}_x(\lambda)$ will be infinite if $\lambda_y \neq 0$ 
for some $y \not\in x+$. Moreover, because of (M) the dual
will be infinite if $\lambda_y <0 $ for some $y$.
Finally, by considering cash balances $K$ that are constant on $x+$ and
using axiom (TI), we see that for finiteness of ${\tilde\pi}_x(\lambda)$
 it is necessary that $\lambda$ be a {\em probability}
on $x+$: $\sum_{y \in x+} \lambda_y = 1$.

One
further property can be deduced: $\inf_\lambda \tilde\pi_x(\lambda)
= \pi_x(0) = 0$, using the duality relation and (Z).
 Thus the dual
valuations $(\tilde\pi_x)_{x \in \sT}$ must satisfy the conditions

\begin{center}%%%%%%%%%%%%%%%%%%%%%%%%%%%%%%%%%%%%%%%%%%%%%
\setlength{\fboxrule}{1pt}
\setlength{\fboxsep}{5mm}
\fbox{
\parbox[c]{0.7\linewidth}{
\nl
(D1) \quad $\tilde\pi_x$ is convex;
\nl
(D2)\quad $\tilde\pi_x(\lambda)$ is only finite if $\lambda$ is a 
probability on $x+$;
\nl
(D3) \quad$\inf_\lambda \tilde\pi_x(\lambda) = 0$.
}
}
\end{center}%%%%%%%%%%%%%%%%%%%%%%%%%%%%%%%%%%%%%%%%%%%%%%%%
\noindent

The recursion for the dual valuations will follow from
the recursive form (\ref{primal_recursion}) of the primal
valuations. To make this explicit, we need to define the
convex duals ${\tilde\pi}_{x,x+1}$ of the one-step valuations
by the usual definition
\[
  {\tilde\pi}_{x,x+1}(\theta,\psi) = \sup_k \{ \;
	\pi_{x,x+1}(k_x,k_{x+1}) - \theta k_x - \psi\cdot k_{x+1} \}.
\]
The analogues of (D1)-(D3) for the dual one-step valuations
will be

\begin{center}%%%%%%%%%%%%%%%%%%%%%%%%%%%%%%%%%%%%%%%%%%%%%
\setlength{\fboxrule}{1pt}
\setlength{\fboxsep}{5mm}
\fbox{
\parbox[c]{0.75\linewidth}{
\nl
(d1) \quad $\tilde\pi_{x,x+1}$ is convex;
\nl
(d2)\quad $\tilde\pi_{x,x+1}(\lambda)$ is only finite if $\lambda$ is 
a probability on $x\cup x+1$;
\nl
(d3) \quad$\inf_\lambda \tilde\pi_{x,x+1}(\lambda) = 0$.
}
}
\end{center}%%%%%%%%%%%%%%%%%%%%%%%%%%%%%%%%%%%%%%%%%%%%%%%%
\noindent
It is easy to see that conditions (c), (m) and (ti) on the
one-step valuations are equivalent to conditions (d1)-(d3)
on their duals.

What then is the dual analogue of the primal recursion
(\ref{primal_recursion})?  The answer is provided by the following
result.

\begin{theorem}\label{T2} 
For all $x \in \sT$ and $\lambda$ a probability on $x+$, we have
\be
  {\tilde\pi}_x(\lambda) = {\tilde\pi}_{x,x+1}(\lambda_x,
{\bar\lambda}_{x+1})
        + \sum_{ z \in x+1} {\bar\lambda}_z
                {\tilde\pi}_z\biggl( \; \frac{ \lambda_{\succeq z} }
                                { {\bar\lambda}_z }\; \biggr)
\label{dual_recursion}
\ee
where $\lambda_{\succeq z}$ denotes the restriction of $\lambda$ to
the set $\{ y: y \succeq z\}$, and
${\bar\lambda}_z \equiv \sum_{y \succeq z} \lambda_y$.
\end{theorem}
\noindent
{\sc Remarks:} (i) Observe that the function 
\[
({\bar\lambda}_z,  \lambda_{\succeq z})
	\mapsto {\bar\lambda}_z
                {\tilde\pi}_z\biggl( \; \frac{ \lambda_{\succeq z} }
                                { {\bar\lambda}_z }\biggr)
	= \sup\{ {\bar\lambda}_z\pi_z(K) - \lambda_{\succeq z} \cdot K \}
\]
is convex.
\nl
(ii) Theorem 4.19 of \cite{CDK2} has a similar flavour to Theorem \ref{T2}.

\noindent
{\sc Proof.} 
Using (\ref{primal_recursion}) and (\ref{convex_dual}), we have
\beast
\pi_x(K) &=& \inf_\lambda\{ \tilde\pi_x(\lambda) + \lambda \cdot K\}
\\
	&=& \pi_x(K_x, \pi_{x+1}(K))
\\
	&=& \inf_{\lambda,\alpha}\{ \; \tpi_{x,x+1}(\lambda_x,\alpha) +
		 \lambda_xK_x +\alpha\cdot\pi_{x+1}(K) \; \}
\\
	&=& \inf_{\lambda,\alpha,\psi}\{ \; \tpi_{x,x+1}(\lambda_x,\alpha) +
		\lambda_xK_x + \alpha\cdot( \tpi_{x+1}(\psi)+\psi\cdot
			K_{[x+1,T]} ) \;\}
\\
	&=& \inf_{\lambda,\alpha,\psi}\bigl[ \; \tpi_{x,x+1}(\lambda_x,\alpha) +
                \lambda_xK_x +  \sum_{z \in x+1} \alpha_z\{ \tpi_z(\psi)
			+ \psi_{\succeq z}\cdot K_{[z,T]} \} \; \bigr]
\\
	&=& \inf_{\lambda,\alpha} \bigl[ \; \tpi_{x,x+1}(\lambda_x,\alpha) +
	   \lambda_xK_x +  \sum_{z \in x+1} \lambda_{\succeq z}\cdot K_{[z,T]}
		+ \sum_{z \in x+1} \alpha_z \tpi_z(\lambda_{\succeq z}/\alpha_z)
			\; \bigr].
\eeast
At the last step, we have made the change of variable
$\lambda_{\succeq z}  \equiv \alpha_z \psi_{\succeq z}$ for $z \in x+1$.
However, the only way that the terms inside the final infimum can be finite
is if the arguments of the dual valuations $\tpi_{x,x+1}$ and $\tpi_z$
are probabilities; and this only happens if $\alpha_z = \bar\lambda_z$ for 
all $z \in x+1$. The conclusion is that
\[
\pi_x(K) = \inf_\lambda \bigl[ \;
	   {\tilde\pi}_{x,x+1}(\lambda_x,
{\bar\lambda}_{x+1})
        + \sum_{ z \in x+1} {\bar\lambda}_z
                {\tilde\pi}_z\biggl( \; \frac{ \lambda_{\succeq z} }
                                { {\bar\lambda}_z }\; \biggr)
	+\lambda\cdot K
	\;\bigr]
\]
and the result is proved.
\hfill$\square$

Notice that if we are given operators $(\tilde\pi_{x,x+1})_{x \in \sT}$
satisfying (d1)-(d3), together with the condition that $\tilde\pi_{x,x+1}
\equiv 0$ for any terminal node $x$, then the corresponding one-step
valuations $(\pi_{x,x+1})_{x \in \sT}$ satisfy (c), (m), (ti),
and $\pi_{x,x+1}(k) = k_x$ for any terminal node $x$.
 By dualising (\ref{dual_recursion}) we quickly arrive
at (\ref{primal_recursion}). Thus we may just as well construct a family
of valuations $\pi_\tau$ satisfying the axioms by starting from a
family $(\tilde\pi_{x,x+1})_{x \in \sT}$ of dual one-step valuations
satisfying (d1)-(d3).

%% file: S4.tex
\section{Examples}\label{S5}
Let us consider some examples which can be analysed
fairly completely in the tree setting. 

\nl
{\bf Example 1: relative entropy.} Suppose given 
some strictly positive
probability distribution $(p_y)_{y \in \sT }$ on $\sT$.
For any $x \in \sT$ we define the dual valuation
$\tilde\pi_x$ evaluated at some probability $\lambda$
on $x+$ to be
\be
\tilde\pi_x(\lambda) = \frac{1}{\gamma} \; \sum_{y \succeq x}\;
		 \lambda_y
\log(\lambda_y \bar p_x/p_y) \equiv 
	h(\,\lambda_{\succeq x}\;|\; p_{\succeq x}/\bar p_x\,) ,
\label{re_dual}
\ee
where $\gamma>0$ is some positive parameter, and
as before $\bar p_x = \sum_{y \succeq x} p_y$. For
other arguments, $\tilde\pi_x$ is infinite.  It is well 
known that the function $\tilde\pi_x$ is convex, and its
concave dual function is easily calculated to be
\be
\pi_x(K) = -\frac{1}{\gamma}\, \log \biggl[ \;\sum_{y \succeq x}\;
	\; \frac{p_y}{\bar p_x} \, e^{-\gamma K_y} \; \biggr],
\label{re_primal}
\ee
equivalently, 
\be
e^{-\gamma \pi_x(K)} = \;\sum_{y \succeq x}\;
        \; \frac{p_y}{\bar p_x} \, e^{-\gamma K_y}.
\label{re_primal2}
\ee
It is now easy to check the axioms (C), (M), (DC), (TI)
of a family of valuation operators, and the axioms (L)
and (CL) will hold by construction when we assemble the 
$(\pi_x)$.

\nl\noindent
{\sc Remark.} From (\ref{re_primal2}) we might conjecture that 
similar examples could be constructed by the recipe
\be
U(\pi_x(K) ) = \sum_{y \succeq x}\; \frac{p_y}{\bar p_x}
                             \; U(K_y)
\label{pi_from_U}
\ee
for some other utility $U$. However, it is not clear that
axioms (TI) and  (C) will be satisfied in general, and indeed 
 the relative entropy example is the only example\footnote{
See also Proposition 2.8 of \cite{Fri4} for a similar result.
}.
%%%%%%%%%%%%%

To see why this is,  if (TI) and (\ref{pi_from_U}) hold, then for
any $a \in \R$ we shall have (assuming with no loss of generality
that $x=0$, and that $p_y>0 \;  \forall y$ to avoid triviality)
\be
U(a + \pi_0(K)) = U( \pi_0(a+K) ) = \sum_{y}\; p_y
                             \; U(a+K_y).
\label{pfU}
\ee
Suppose we choose some non-constant $K$ for which $\pi_0(K) = 0$, and
perturb $K_y$ to $K_y + t v_y$; differentiating (\ref{pfU}) with
respect to $t$ at $t=0$ yields
\[
U'(a )\sum_y v_y \frac{\partial \pi_0(K)}{\partial K_y}
	= \sum_y p_y v_y U'(a+K_y)
\]
implying that 
\[
  \sum_y v_y \frac{\partial \pi_0(K)}{\partial K_y}
	= \sum_y p_y v_y \frac{U'(a+K_y)}{U'(a)}
\]
does not depend on $a$. Now select some $y$ for which $K_y
=\delta >0$, and make $v_y=1$, $v_z = 0$ for $z \neq y$.
We conclude that $U'(a+\delta)/U'(a)$ does not depend on 
$a$. Monotonicity of $U'$ and the fact that $\delta>0$
can be chosen arbitrarily imply that $U'$ is exponential.

\nl\nl
%%%%%%%%%%%%%%%%%%%%%%%%%%%%%%%%%%%%%%%%%%%%%%%%%%%%%%%%%%%%%%
{\bf Example 2.} This example is really a family of examples,
built from the simple observation that if we have some collection
$(\pi^\theta_{x,x+1})_{x \in \sT, \theta \in \Theta}$ of 
one-step valuation operators, such that for each $\theta$ the
family $(\pi^\theta_{x,x+1})_{x \in \sT}$ satisfies the 
axioms (c), (m) and (ti), then the one-step valuation operators
defined by 
\be
	\pi_x(k) \equiv \inf_{\theta} \pi^\theta_{x,x+1}(k)
\label{eg2}
\ee
again satisfy (c), (m) and (ti).

One simple example of this form could be constructed as 
follows. Suppose that for each $x \in \sT$ we have some 
probability distribution $\alpha(x)$ on the immediate
descendents $x+1$, and now we define
\[
   \pi_{x,x+1}(k) = \min\{ k_x, \sum_{y \in x+1} \alpha(x)_y k_y\}.
\]
The recursion (\ref{primal_recursion}) is now just the
Bellman equation of dynamic programming, and the
 value of $\pi_x(K)$ is the `worst stopping' value
of the Markov decision process, where $\alpha(x)$ gives 
the distribution of moves down to $x+1$ from $x$ if it is
decided not to stop at $x$. It is easy to extend this example
to the situation where a finite collection of possible distributions
$\alpha^i(x)$ is considered at each vertex $x$, and the valuation
operator gives the `worst worst stopping' value!

Several of the examples of \cite{CDK2} are of this form, and we
make no further remark on them. However, one feature that is
 noteworthy is 
the following. If we make the one-step valuation operators as
infima of some sequence of linear functionals, 
\[
	\pi_{x,x+1}(k) = \min_{j} \alpha^j \cdot k
\]
then the valuations constructed are coherent, and 
the dual one-step valuation operators $\tilde\pi_x$ are
\beast
\tilde\pi_{x,x+1}(\lambda) &=& 0 \quad\hbox{\rm if $\lambda \in \co(
                       \{ \alpha^j \} )$}
\\
&=& \infty \quad\hbox{\rm  otherwise,}
\eeast
where $\co(A)$ denotes the convex hull of the set $A$.
Looking at the recursive form (\ref{dual_recursion}) of the 
dual valuation
operators, we see that these too take only the values 0 and $\infty$.
The set of $\lambda$ for which $\tilde\pi_0(\lambda)$ is finite is a 
{\em mutiplicatively stable} set.

\nl\nl
%%%%%%%%%%%%%%%%%%%%%%%%%%%%%%%%%%%%%%%%%%%%%%%%%%%%%%%%%%%%%%
{\bf Example 3.} Families of one-step valuations $\pi_{x,x+1}$
can be constructed via the notion of a utility-indifference 
price for a single-period problem.
 In more detail, given some probability distribution
$p_y$ over $ x \cup x+1$, we define $\pi_{x,x+1}(k)$ to be 
that value $b$ such that
\begin{equation}
   U(x_0) = \sum_{y \in x \cup x+1} p_y U(x_0 + k_y -b)
\label{uip1}
\end{equation}
where $U$ is some strictly increasing utility function, and
$x_0$ is some reference wealth level. The properties (m) and
(ti) are immediate, and (c) is a simple deduction.

It is unfortunately the case that there are few examples
where the utility-indifference price  for a single-period 
problem can be computed in closed form, and the dual
valuation is similarly elusive. Some progress can be made
however. Dropping the subscripts, the calculation of the 
dual valuation requires us to find
\[
	\tilde\pi(\lambda) = \sup_k\{ \pi(k) - \lambda\cdot
		k \}
\]
and the optimisation here can be considered as the optimisation
\begin{gather}
	\sup_{b,k} \; b-\lambda\cdot k
\\
\hbox{\rm subject to} \quad U(x_0) = \sum p_y U(x_0 + k_y -b).
\end{gather}
The Lagrangian form of the problem 
\[
  \sup_{b,k} \bigl[ b - \lambda\cdot k
            +\theta ( \sum p_y U(x_0 + k_y -b) -U(x_0) ) \bigr]
\]
leads to the first-order conditions
\beast
1 &=& \theta \sum p_yU'(x_0 + k_y -b),
\\
\lambda_y &=& \theta p_yU'(x_0 + k_y -b),
\eeast
so (with $I \equiv (U')^{-1}$) we get
\[
	x_0 + k_y -b = I\bigl(\frac{\lambda_y}{\theta p_y}\bigr),
\]
from which we see that $\theta$ is determined via
\[
	 \sum p_y U\biggl(I\bigl(\frac{\lambda_y}{\theta p_y}\bigr)
		\biggr) = U(x_0).
\]
The final expression
\[
	\tilde\pi(\lambda) = x_0 
    - \sum \lambda_yI\biggl(\frac{\lambda_y}{\theta p_y}\biggr)
\]
simplifies in the case of CRRA $U(x) = x^{1-R}/(1-R)$ to 
\[
  \tilde \pi(\lambda) = x_0 - x_0^R\Biggl(
     \sum p_y^{1/R} \lambda_y^{1-1/R} \Biggr)^{R/(R-1)}.
\]

\nl
{\sc Remarks.} This example constructs a family of valuations
from one-step valuations which can be interpreted as utility-indifference
prices.   Utility-indifference pricing is a popular method
for pricing in incomplete markets (at least for the pricing of a
European-style contingent claims\footnote{
The study \cite{Cheridito_Kupper} tackles 
utility-indifference pricing in the context
of risk measurement.
}); how does it relate to what we
are doing here? Suppose we take  a cash-balance process which is 
zero at all non-terminal nodes. We can for each $x \in \sT$ compute
the utility-indifference price starting from node $x$, but is this
recipe consistent with the axioms we have set down here? Of course, 
the utility-indifference prices can only be computed (at least
in the first place) for terminal contingent claims; we do not 
know to start with how we might price more complicated cash balances
from utility indifference. Nevertheless, we can already say 
that in general such utility-indifference prices do {\em not}
satisfy the axioms we are considering here.

To see why, consider a two-period trinomial tree, with each of the
nine terminal nodes having equal probability. We assume that the
agent has a CRRA utility, and 
 will receive a baseline payment of 2 at each terminal node,.
He now tries to value a contingent claim that pays $y_\omega$
at $\omega$. It is an easy matter to calculate the 
utility-indifference price at the root node, and at each of the
time-1 nodes.  Now if these valuations are to satisfy (DC), if
we consider different $y$ which all have the same time-1
utility-indifference prices, then they must have the same time-0
utility-indifference price; numerical examples demonstrate that
they do {\em not}, 
%  /home/chris/WORK/GRADS/ARNAUD/pricing/uip/run3
and so utility-indifference pricing does not
satisfy the axioms we have given for valuations.

%% file: S5.tex
\section{Spreading and evolution of risk.}%%%%%%%%%%%%%%%%%%%%%%%%%%%%%%
\label{S4} Let us consider the situation of a firm which 
consists of $J$ subsidiaries, possibly in different countries,
or subject to different regulatory controls. We let the
valuations $(\pi_\tau^j)_{\tau \in \sO}$ determine
the regulatory requirements of subsidiary $j$, $j = 1,\ldots, J$.
If (at $\tau$)
 subsidiary $i$ wishes to accept the cash balance process $K$, 
then regulation requires that subsidiary to reserve $-\pi_\tau^i(K)$.
\\
{\sc Remark.}
This implicitly supposes that the subsidiary faces a zero cash 
balance process, and that it will only incur regulatory capital
requirements if it changes this. This contrasts with \cite{BEK},
\cite{JST1}, where it is supposed that the subsidiary has already
entered into some commitments, which have involved the acceptance
of cash balance $K^*$, say; their risk-sharing results then depend on
the $K^*$ for each subsidiary.
 In our treatment, if the subsidiary is already committed
to $K^*$, then we introduce the valuations $\pi_\tau^*(K) 
\equiv \pi_\tau(K + K^*) - \pi_\tau(K^*)$ (it can be checked that
these {\em are} valuations, satisfying the axioms (C), (L), (CL),
(M), (DC), (TI), (Z)), and proceed with these. This is notationally
 simpler; the results do indeed depend on the subsidiaries'
prior commitments, though this dependence does not appear explicitly.
The interested reader is invited to make appropriate notational
changes to express the dependence on $K^*$ explicitly if desired.
\nl
\nl
To reduce its risk, subsidiary $i$ could approach another subsidiary $j$
and get them to take from  $i$ the cash-balance process $K^j$ in 
return for the regulatory capital $-\pi_\tau^j(K^j)$.  Subsidiary $i$
is free to enter into such agreements with all the other subsidiaries,
and will do so in such a way as to minimise the regulatory capital
required. Taking into account the possibilities of risk transfer,
subsidiary $i$ will need to reserve $-\Pi_\tau^i(K)$ instead of
$\pi_\tau^i(K)$, where
\bea
\Pi_\tau^i(K) &=&  \sup\{ \pi_\tau^i(K - \sum_{j \neq i} K^j 
		+\sum_{j \neq i} \pi_\tau^j(K^j)I_{[\tau,T]}) \}
\nonumber
\\
&=& \sup\{ \pi_\tau^i(K - \sum_{j \neq i} K^j) + \sum_{j \neq i} \pi_\tau^j(K^j) \}
\nonumber
\\
&=& \sup\{ \sum_j \pi_\tau^j(K^j) : \sum_j K^j = K \}.
\label{rs1}
\eea
Notice that this is {\em independent of the choice of subsidiary}, so we 
write simply $\Pi_\tau$ for $\Pi_\tau^i$. Moreover, we may have that 
$\Pi_\tau(0)>0$, so we define
\be
	\opi_\tau(K) = \Pi_\tau(K) - \Pi_\tau(0),
\label{Pi_def}
\ee
so as to have the property (Z) for the operators 
$\opi_\tau$. The quantity $\Pi_x(0)$ can be interpreted as the
{\em value of risk-sharing} at vertex $x$.  We call the family
$(\opi_\tau)_{\tau \in \sO}$ of valuations the {\em risk-sharing}
valuations, though it is not clear as yet that we may 
refer to them as such, since we do not know that they satisfy
the axioms for a family of valuations. That is the 
task of the following result.

\begin{theorem}\label{thm3}
 The risk-sharing valuations $(\opi_\tau)_{\tau \in \sO}$
satisfy the axioms (C), (L), (CL), (M), (DC), and (TI) of the 
component valuations $(\pi_\tau^j)_{\tau \in \sO}$, $j = 1,\ldots, J$.
\end{theorem}

\nl\nl
{\sc Proof.}
Properties (C), (L), (CL), (M), and (TI) are straightforward
to verify; only the property
(DC) is not immediately obvious. To establish this, we have on the one hand
\bea
\Pi_x(K)
	&=&
	 \sup\{ \sum_j \pi_x^j(K^j) : \sum_jK^j = K\}
\nonumber
\\
&=&  \sup\{ \sum_j \pi_x^j(K^jI_{[x,\tau)}
		+ \pi_\tau^j(K^j)I_{[\tau,T]}  ) :  \sum_jK^j = K\}
\nonumber
\\
&=& \sup\{ \sum_j \pi_x^j(K^jI_{[x,\tau)}
                + a_j I_{[\tau,T]}  ) :  \sum_jK^jI_{[x,\tau)} = KI_{[x,\tau)},
\nonumber
\\
&& \quad\quad\quad\quad\quad\quad\quad\quad\quad\quad\quad\quad
			\sum_j a_j \leq \Pi_\tau(K) \}
\label{rs2}
\eea
and on the other hand we have
\bea
\Pi_x(K_{[x,\tau)} +\opi_\tau(K)I_{[\tau,T]})
        &=& \sup\{ \sum_j \pi_x^j(K^j) : \sum_jK^j = K_{[x,\tau)}
		 +\opi_\tau(K)I_{[\tau,T]}  \}
\nonumber
\\
&=& \sup\{ \sum_j \pi_x^j(K^jI_{[x,\tau)}
                + \pi_\tau^j(K^j)I_{[\tau,T]}  ) :  
\nonumber
\\
&& \quad\quad\quad\quad\quad\quad\quad\quad\quad\quad\quad\quad
\sum_jK^j = K_{[x,\tau)}
                 +\opi_\tau(K)I_{[\tau,T]}  \}
\nonumber
\\
&=& \sup\{ \sum_j \pi_x^j(K^jI_{[x,\tau)}
 	+a_jI_{[\tau,T]}  ) :  \sum_jK^jI_{[x,\tau)} = KI_{[x,\tau)},
\nonumber
\\
&& \quad\quad\quad\quad\quad\quad\quad\quad\quad\quad\quad\quad
		\sum_j a_j \leq \Pi_\tau(\opi_\tau(K)I_{[\tau,T]}) \}
\label{rs3}
\eea
But $\Pi_\tau(\opi_\tau(K)I_{[\tau,T]}) = \Pi_\tau(0)+\opi_\tau(K)
=\Pi_\tau(K)$ and comparing (\ref{rs2}) and (\ref{rs3}) we see that
$\Pi_x(K) = \Pi_x(K_{[x,\tau)} +\opi_\tau(K)I_{[\tau,T]})$, equivalently,
$\opi_x(K) = \opi_x(K_{[x,\tau)} +\opi_\tau(K)I_{[\tau,T]})$, as required.

\eoproof
%%==================================================================

\nl\nl
There is a simple interpretation of risk-sharing in terms of 
the duals. Indeed, from (\ref{rs1}) we have that
\beast
\tilde\Pi_x(\lambda) &=& \sup_{ (K^j)} \{ \sum_j \pi_x^j(K^j) -\lambda\cdot
			\sum_j K^j \}
\\
&=&  \sum_j \tilde\pi_x^j(\lambda),
\eeast
so the effect of risk-sharing is simply to {\em add the dual valuations}.

%%%%%%%%%%%%%%%%%%%%%%%%%%%%%%%%%%%%%%%%%%%%%%%%%%%%%%%%%%%%%%%%%%%%%
\subsection{Optimal risk transfer in the relative entropy example.}
\label{ORS_RE}
If each of the $J$ subsidiaries has valuations of the relative
entropy form (recall (\ref{re_primal})):
\be
e^{-\gamma_j \pi_x^j(K)} = \;\sum_{y \succeq x}\;
        \; \frac{p^j_y}{\bar p^j_x} \, e^{-\gamma_j K_y},
\label{re_agent_j}
\ee
how do they  combine under risk sharing? For the moment,
let us fix a particular $x \in \sT$ and consider how things work
from that node.  We shall write $\tilde p^j_y \equiv p^j_y/\bar p_x^j$
for brevity, and shall define
\be
	\Gamma \equiv \bigl( \sum_j \gamma_j^{-1} \bigr)^{-1}.
\label{Gamma_def}
\ee
The dual valuations are given by (see (\ref{re_dual}) )
\[
\tilde\pi^j_x(\lambda) = \frac{1}{\gamma_j} \;
   \sum_{y \succeq x}\; \lambda_y
           \log(\lambda_y /\tilde p_y^j),
\]
so the risk-sharing result  gives us 
\beast
	\tilde \Pi_x(\lambda) &=& \sum_j \tilde\pi^j_x(\lambda)
\\
	&=& \sum_j\frac{1}{\gamma_j} \;
   \sum_{y \succeq x}\; \lambda_y
           \log(\lambda_y /\tilde p_y^j)
\\
	&=& \frac{1}{\Gamma} \,\biggl\lbrace \; 
	 \sum_{y \succeq x}\; \lambda_y
	   \log \lambda_y  - \sum_j\frac{\Gamma}{\gamma_j} \;
   \sum_{y \succeq x}\; \lambda_y \log \tilde p_y^j\; \biggr\rbrace
\\
	&=& \frac{1}{\Gamma} \,\sum_{y \succeq x}\; \lambda_y
		\log(\lambda_y/P_y) 
%-\frac{1}{\Gamma} \, \log\biggl\lbrace
                %\sum_{y \succeq x}\; 
   %\exp( \sum_j \frac{\Gamma}{\gamma_j}\,\log(\tilde p_y^j) )
%\prod_i (p^i_y)^{\Gamma/\gamma_i}
                        %\, \biggr\rbrace
 + \frac{1}{\Gamma} \, \log A,
\eeast
where we define the probability $P$ on $x+$ and the constant $A$ by
\bea
	P_y &\equiv& A \;  \prod_i (\tilde p^i_y)^{\Gamma/\gamma_i}
        	 \equiv \frac{  \prod_i (\tilde p^i_y)^{\Gamma/\gamma_i}}
		{ \sum_{z \succeq x} \prod_i (\tilde p^i_z)^{\Gamma/\gamma_i}}.
\label{P_def}
\\
   A &\equiv& \biggl( \; \sum_{z \succeq x} \prod_i (\tilde 
                              p^i_z)^{\Gamma/\gamma_i}
		\; \biggr)^{-1}
\label{Adef}
\eea
From this we see that 
\beast
\Pi_x(0)  &=& \frac{1}{\Gamma} \, \log A
\\
&=&
  -\frac{1}{\Gamma} \, \log\biggl\lbrace
                \sum_{y \succeq x}\;
\prod_i (\tilde p^i_y)^{\Gamma/\gamma_i}
                        \, \biggr\rbrace
\\
&=&	
-\frac{1}{\Gamma} \, \log\biggl\lbrace
                \sum_{y \succeq x}\; \exp( \sum_j
                \frac{\Gamma}{\gamma_j}\,\log(\tilde p_y^j) )
                        \, \biggr\rbrace
\\
	&\geq& 0,
\eeast
by Jensen's inequality, with equality if and only if all the agents
have the same $p^j_y$. Thus we see that the aggregated dual valuations
$\tilde\opi_x$ have the {\em same} relative-entropy form as
the individual dual valuations, with explicit expressions 
(\ref{Gamma_def}) for the combined coefficient of
absolute risk aversion $\Gamma$ and  (\ref{P_def}) for the combined
distribution of the probability down the tree.

How does the risk sharing work out in this example?  The  maximisation
(\ref{rs1}) of $\sum_j \pi_x^j(K^j)$ can be computed, leading to 
the conclusion that
\bea
   K_y^j &=& \frac{\Gamma}{\gamma_j} \, K_y
	+\, \biggl\lbrace \; \frac{1}{\gamma_j}\log \tilde p_y^j
	-\frac{\Gamma}{\gamma_j}\biggl(
		\sum_i \frac{1}{\gamma_i}\log \tilde p_y^i
			\biggr) \; \biggr\rbrace.
\label{Krs}
\\
&=& \frac{\Gamma}{\gamma_j} \, K_y
    +\frac{1}{\gamma_j}\log( \tilde p_y^j / P_y)
	+\frac{\Gamma}{\gamma_j} \Pi_x(0)
\label{Krs2}
\eea
This provides a nice interpretation of the way that the 
cash balance $K$ gets shared. At each node $y$, the cash balance
$K_y$ at the node gets split proportionally between the subsidiaries
(`linear risk sharing' as in \cite{borch}),
 and there are a further  two terms, 
one relating to the ratio of subsidiary $j$'s probability of the 
node $y$ and the aggregated probability $P_y$, and the other 
proportional to $\Pi_x(0)$.

%%%%%%%%%%%%%%%%%%%%%%%%%%%%%%%%%%%%%%%%%%%%%%%%%%%%%%%%%%%%%%%%%%%%%
\subsection{Dynamic stability of the risk-sharing solution.}
\label{DSRS}
When computing the value $\Pi_0(0)$ of risk-sharing at time 0, 
the subsidiaries find themselves solving the optimisation problem
\[
\sup\{ \sum_j \pi_0^j(K^j) : \sum_j K^j = 0 \}.
\]
Casting the problem in Lagrangian form
\[
  \sup \{ \sum_j \, [ \, \pi_0^j(K^j) - p \cdot K^j \, ] \},
\]
it is easy to see that at an optimal solution we shall have
that all subsidiaries' marginal valuations of cash balances will
coincide:
\be
	\nabla \pi_0^j(K^j) = p.
\label{eqm_prices}
\ee
Suppose that at time 0 they adopt the optimal cash balance
processes $K^j$ obtained in this way; as time passes, {\em will they
still be satisfied with the $K^j$ they first agreed to?} It would
be disturbing if we reached some vertex $x$ in the tree where the
subsidiaries would wish to renegotiate the deals that they had committed to
at time 0.  However, it turns out that {\em this does not happen}:
and it is the condition (DC) and the chain rule which guarantees this.

If $x$ is some vertex in the tree, and we let $\tau = \tau_x$
(recall (\ref{tau_x_def})), then using (DC) we have
\[
\pi_0^j(K) = \pi_0^j(KI_{[0,\tau)}+ \pi_\tau^j(K)I_{[\tau,T]} )
\]
and differentiating both sides with respect to $K_y$, where
$y \succeq x$, gives us (by the chain rule)
\[
p_y = \frac{\partial \pi_0^j}{\partial K_y} (K^j)
   = \frac{\partial \pi_0^j}{\partial K_x}
         (K^jI_{[0,\tau)}+ \pi^j_\tau(K^j)I_{[\tau,T]} )
        \frac{\partial \pi_x^j}{\partial K_y}( K^j ).
\]
Accordingly, in view of (\ref{eqm_prices}), we have for each $j$
that there exists a constant $b_j$ such that for all $y \succeq x$
\[
        \frac{\partial \pi_x^j}{\partial K_y}(K^j)
                 = b_j p_y,
\]
and so at vertex $x$ the remaining allocations (cash balances)
 $K^jI_{[x,T]}$ still constitute a competitive equilibrium;
there are no mutually beneficial trades available to the agents
at vertex $x$.

\noindent
{\sc Remarks.} We could have discussed this dynamic stability in 
terms of competitive equilibria. Indeed, if we were to write 
$U^j(K)$ in place of $\pi^j_0(K)$, then the concave increasing
functions $U^j$ can serve
as the utilities of different agents, defined over bundles of 
goods, where cash balances at different vertices are interpreted
as different goods. We are now in the realm of finding an
equilibrium allocation, which is a standard part of
 microeconomic theory; see, for example, \cite{MWG}.  However,
although the mathematics is exactly that of finding an equilibrium
in a pure exchange economy, such an analogy is economically impure;
here, we have been interpreting the $\pi^j_x$ as some sort of
 {\em price}, not as a utility.
We allow (for example) the values $\pi_x^j(K)$ into the {\em arguments}
of the functions  $\pi_0^j$. It seems to us that the
link is tenuous, and we desist from pushing the analogy too far.

%%%%%%%%%%%%%%%%%%%%%%%%%%%%%%%%%%%%%%%%%%%%%%%%%%%%%%%%%%%%%%%%%%%%%
\subsection{Spreading risk by access to a market.}\label{access}%%%%%
Suppose a firm with valuations  $(\pi_x)_{x \in \sT}$
is allowed access to a market; how will it act, and how does its valuation
of cash balances change? The discussion is similar to that of risk sharing
among subsidiaries,
but sufficiently different to require a separate treatment\footnote{
The effect of access to a market in the context of dynamic valuation
of terminal contingent claims is addressed also in \cite{Kloeppel_Schweizer}.
}.

We represent access to the market in the following way. At each stopping
time $\tau$, the firm may change a given cash balance process $K$ to $K+K'$
for any $K' \in G_\tau$, where $G_\tau$ denotes the gains-from-trade
cash balance processes which could be achieved by trading in 
the market starting with zero wealth at time $\tau$.
Concerning the $G_\tau$ we shall assume\footnote{
Recall the definition (\ref{tau_x_def}) of $\tau_x$.
}
 that
\nl
(c-m) each $G_\tau$ is convex;
\nl
(l-m) for each $x \in \lb\tau\rb$,
\[
	G_{\tau_x} = \{ KI_{[x,T]}: K \in G_\tau \};
\]
\nl
(dc-m) for each $\tau \leq \sigma \in \sO$  if $K^\sigma$ denotes\footnote{
Formally, $K^\sigma_z = K_y$ if $z \succeq y \in \lb\sigma\rb$; $=K_z$ otherwise.
} the cash balance
process $K$ stopped at $\sigma$, we have
\[
	 G_{\tau} = \{ K^\sigma+ K': K \in G_\tau, K' \in G_\sigma\}
\]

\nl\nl
Now the cash balance valuation given access to this market will be
via
\be
	\Pi_x(K) \equiv \sup\{ \pi_x(K+K'): K' \in G_x \}.
\label{access1}
\ee
Once again, there is no guarantee that $\Pi_x(0) = 0$, but if we define
\be
	\opi_x(K) \equiv \Pi_x(K) - \Pi_x(0),
\ee
then the operators $(\opi_x)_{x \in \sT}$ do have this property
(Z). As in the case of risk-sharing, the quantity
$\Pi_x(0)$ is the value to the agent of being granted access to the 
market at vertex $x$.

\begin{theorem}\label{thm4}
 The  valuations $(\opi_\tau)_{\tau \in \sO}$
satisfy the axioms (C), (L), (CL), (M), (DC), and (TI).
\end{theorem}
\nl\nl
{\sc Proof.}
As before, all of the properties except
for (DC) are obvious. To prove (DC), we use properties (DC) for the $\pi_x$ 
and (dc-m) to develop
\bea
\Pi_x(K) &=& \sup\{ \pi_x(K+K') : K' \in G_x \}
\nonumber
\\
	&=& \sup\{ \pi_x( (K+K')I_{[x,\sigma)} + 
		\pi_\sigma(K+K')I_{[\sigma,T]}):  K' \in G_x\}
\nonumber
\\
	&=& \sup\{ \pi_x( (K+K')I_{[x,\sigma)} +
            \{ \Pi_\sigma(K) + K'_\sigma \} I_{[\sigma,T]}):  K' \in G_x\}.
\label{am1}
\eea
On the other hand, 
\bea
\Pi_x(KI_{[x,\sigma)}+ \opi_\sigma(K)I_{[\sigma,T]} ) 
    &=&
   \sup\{ \pi_x( KI_{[x,\sigma)}+ \opi_\sigma(K)I_{[\sigma,T]}+K'):K'\in G_x\}
\nonumber
\\
&=& \sup\{ \pi_x( (K+K')I_{[x,\sigma)}+ 
	\opi_\sigma(K)I_{[\sigma,T]}+K'I_{[\sigma,T]}):K'\in G_x\}
\nonumber
\\
&=&\sup\{ \pi_x(\; (K+K')I_{[x,\sigma)}+
        \opi_\sigma(K)I_{[\sigma,T]}+
\nonumber
\\
&&\quad\quad\quad\quad\quad\quad\quad\quad
		+\pi_\sigma( K'I_{[\sigma,T]})I_{[\sigma,T]}\;) :K'\in G_x\}
\nonumber
\\
&=&\sup\{ \pi_x(\; (K+K')I_{[x,\sigma)}+
        \opi_\sigma(K)I_{[\sigma,T]}+
\nonumber
\\
&&\quad\quad\quad\quad\quad\quad\quad\quad
                + (\Pi_\sigma(0) + K'_\sigma)I_{[\sigma,T]}\;) :K'\in G_x\}
\label{am2}
\eea
since when we maximise $\pi_\sigma( K'I_{[\sigma,T]})$ over $K'$ we
get $\Pi_\sigma(0) + K'_\sigma$. Comparing (\ref{am1}) and (\ref{am2})
establishes property (DC) for the operators $( \opi_\tau)_{\tau \in \sT}$.
\eoproof

%% file: conclusions.tex
\bigskip
\section{Conclusions.}\label{conclusions}%%%%%%%%%%%%%%%%%%%%%%%%%%%%%%%%%%%
This paper has approached the problem of convex risk measurement in 
a dynamic setting from a slightly unconventional starting point; instead of 
trying to work with acceptance sets, we begin with valuations
satisfying certain axioms which seem to us to be natural.  Our notion
of preference does {\em not} reduce to  a simple valuation of all the
proceeds of the cashflow collected at the end, but genuinely accounts
for the (obvious) fact that you would prefer to have \$1M today
rather than the value of \$1M  invested at riskless rate
 in five years from now.

In the simplest situation, where the sample-space is finite, we show
how a family of pricing operators obeying our axioms can be decomposed into
(and reconstructed from) a family of one-period pricing operators which
are much easier to grasp. There is a corresponding decomposition of the
dual pricing functions.  

Allowing  a firm to spread risk among a number of subsidiaries
 leads to  risk-sharing solutions;
the firm derives benefit from risk sharing,
and, remarkably, the  risk-sharing valuations which arise 
satisfy exactly the same set of axioms satisfied by the initial
valuations.

We have seen also that the risk sharing that arises will
be stable over time; if at time 0 the firm chooses
how to spread risk among its subsidiaries,
  then no matter how the 
world evolves, at all later times it will continue to be
satisfied with the cash balances that it originally selected.

We study also what happens when a firm is allowed access to
a financial market. Assuming some natural properties of the market, 
the conclusions are similar to the risk-sharing problem; the firm
derives a fixed benefit from being allowed access to the market, but
beyond that it values cash balance processes according to modified
valuations which satisfy the same axioms.

%% file: appendix.tex
\section{Appendix.}
\nl
{\bf Proof of Theorem 1.}
Let us consider the filtered probability
space $(\Omega, ( {\cal F}_t )_{t \geq 0}, P)$.
For any \( T > 0 \), the map \( A \mapsto \pi_{0T} (I_A) \) defines a
non-negative measure on the \( \sigma \)-field \( \mathcal{F}_T \), from the
linearity, positivity and continuity properties of our pricing operator. 
\newline
\newline
This measure is moreover absolutely continuous with respect to \( {P} \) in
view of (A2). Hence by the Radon-Nikodym theorem, 
for any $T>0$, there exists a non-negative \( \mathcal{F}_T
\)-measurable random variable \( \zeta_T \) such that
$$\pi_{0T}(Y) = {E}[\zeta_T Y]$$
for all \( Y \in L^{\infty}(\mathcal{F}_T) \). 
\newline
\newline
Moreover, (A2) implies
that \( {P}[\zeta_T > 0] = 1 \). 
\newline
\newline
We finally use the consistency
condition (A3) as follows. Let \( Y \in L^{\infty}(\mathcal{F}_T) \), then by definition, \( \pi_{tT}(Y) \in L^{\infty}(\mathcal{F}_t) \). For any \( X \in L^{\infty}(\mathcal{F}_t) \),
\beast
      \pi_{0t}(X \pi_{tT}(Y)) &=& {E} [X \zeta_t \pi_{tT}(Y)] 
\\
    &=& \pi_{0T}(XY)
\\
    &=& {E}[XY \zeta_T].
\eeast
Since \( X \in L^{\infty}(\mathcal{F}_t) \) is arbitrary, we deduce that
$$\pi_{tT}(Y) = \frac{1}{\zeta_t} {E}_t [\zeta_T Y]$$
which shows that the pricing operators \( \pi_{st} \) are actually given by the
risk-neutral pricing recipe \eqref{price} described in Theorem 1, with the state-price density process \( \zeta \).
\newline
\newline
The state-price density process is often thought of as the product of the 
discount factor $ \exp \left( - \int_{0}^{t} r_s \textrm{d}s \right) $ and 
the change-of-measure martingale.